# Self-Healing Small-Scale Swimmers

*Emil Karshalev[1] †, Cristian Silva-Lopez[1] †, Kyle Chan[2], Jieming Yan[1], Elodie Sandraz[1], Mathieu Gallot[1], Amir Nourhani[1], Javier Garay[2] and Joseph Wang[1] \**

[1] Department of NanoEngineering, University of California San Diego, La Jolla, CA 92093, USA.
[2] Department of Mechanical and Aerospace Engineering, University of California San Diego, La Jolla, CA 92093, USA.

*Corresponding author. Email: josephwang@eng.uscd.edu.

**Keywords:** Self-Healing robotics, Magnetic, Composite materials, Self-propelled, Autonomous robotics

**Abstract**

Herein, self-healing small-scale swimmers capable of autonomous propulsion and 'on-the-fly' structural recovery are described. The new strategy instantaneously restores the functionality of the swimmer after it has suffered severe damage. Incorporation of magnetic microparticles in strips along with the printed functional body layers (consisting of conductive carbon, low-density hydrophobic polymer and catalytically active metal) results in rapid reorientation and reattachment of the moving damaged catalytic tail to its complimentary broken static body piece. Such magnetic alignment and attraction restores the original swimmer structure and propulsion behavior, independent of user input, displaying healing efficiencies as high as 88%. Modeling of the magnetic fields and simulations of various swimmer configurations are used to study the magnetic force field distribution around the swimmers. The influence of the damage position and pattern of multiple magnetic healing strips is examined and their influence upon healing efficiency is compared. Owing to its versatility, fast recovery, simplicity, and efficiency, the new 'on-the-fly' self-healing strategy represents an important step towards the development of new classes of robots that can regain their functionality in situations of extreme mechanical damage where repair is not possible or challenging.

Small- and micro-scale robots have recently featured in numerous studies showing considerable potential for diverse environmental and security remediation, sensing applications or biomedical drug-delivery and surgery applications.[1-8] Often these tiny robots are released into harsh environments where multiple hazards can lead to cracks, rips, tears and breakage of the robots, resulting in catastrophic failure and cessation of the robot motion and operation. Large, hard, metal-based robots are composed of a large array of replaceable parts and do not suffer the limitations of soft polymeric or hydrogel robots. The latter are vulnerable to damage because of low tear strength and propensity for crack growth, despite of their softness and flexibility.[9-11] Furthermore, optimal self-healing strategies require healing to occur autonomously without user input or additional external triggers, which is in contradiction with traditional temperature or light-based chemical healing approaches. Additionally, most robots can experience damage in the same place more than once, requiring new healing strategies for extending a robot's lifetime for such repetitive damage under dynamically changing conditions.

Material innovations and engineering ingenuity have been combined to yield specialized robotics systems capable of responding to mechanical damage. One strategy features a soft robot skin that can sustain large mechanical deformation due to the formation of liquid metal frameworks within a silicone elastomer but once the material is damaged and depleted there is no way to recombine it.[12] More closely resembling a robotic appendage, Acome *et al.* designed hydraulically amplified self-healing electrostatic (HASEL) actuators which can easily regain their actuation performance even after 50 dielectric breakdown cycles.[13] The nature of the liquid dielectric prevents permanent damage through dielectric breakdown but comes at the price of diminished performance. Another self-healing strategy for soft robotics utilizing a robotic gripper relies on the thermoreversible Diels-Alder reaction to close cracks and punctures; yet it is not autonomous and requires the application of an external stimulus (temperature) over extended periods (minutes to days).[9] Smaller scale robotic systems utilized magnetic interactions to assemble robots into various shapes with a high degree of control.[14,15] However, these strategies require extensive human involvement with continuous manipulation of multiple magnetic coils and fluidic channel flows, are not explicitly for self-healing and are only concerned with assembly on very small length scales. Another major concern regarding existing self-healing strategies (based on different chemical and physical principles) is the inability to heal under different scenarios. For example, strategies involving chemical bonding and capsule based systems are easily affected by ambient conditions, limiting the healing behavior in single site and making them not suitable for healing under harsh environments.[17] Additionally, these strategies experience limitations at shorter timescales as the healing must occur 'on-the-fly', or over many damage cycles in the same location.

Herein, we present a robust and efficient autonomous 'on-the-fly' self-healing approach for small-scale untethered artificial chemically-powered swimmers. Since small-scale robots are designed for active operation, our swimmers are autonomously propelled, highlighting further their unique ability to repair ,'on-the-fly'. Movement is achieved by the catalytic decomposition of a peroxide fuel at the catalytic Pt surface that generates an oxygen bubble thrust.[18-20] The self-healing swimmer (SHS) utilizes strong magnetic interactions to recover its normal swimming function. Such new healing strategy is autonomous and can recover the structure of the swimmer instantaneously even after being broken catastrophically into multiple pieces, while restoring the original propulsion behavior. The composite layered structure of the swimmer has been optimized to feature distinct material layers for different functions: rigidity, propulsion, scaffolding and magnetic attraction. The built-in magnetic torque thus aligns and attracts the damaged pieces without user input or additional external triggers. Additionally, the geometrical design of the swimmer, particularly the alignment of the magnetic layer, enables a high healing efficiency, with 88% recovery efficiency of the initial structure.

Magnetic properties are attractive for such healing behavior as they are not readily inhibited by environmental conditions. While magnetic self-healing strategies using iron-oxide particles have been developed before they have not been applied to robotics or as a standalone self-healing mechanism but usually in connection to hydrogel or polymeric healing materials.[21,22] However, the incorporation of strongly magnetic $Nd_2Fe_{14}B$ microparticles has been reported as an efficient strategy to heal printed electronic devices.[23] The self-healing magnetic layer is incorporated in the form of a strip, or several strips, within the main body of the catalytic swimmer. After suffering critical damage, the SHS is split into multiple pieces, some of which are static and some (containing the catalytic tail) are active and moving. The active piece continues to propel itself while the magnetic attraction between all separated pieces containing the magnetic strip, enables strong and lasting reattachment and recovery of the initial structure and of the original propulsion behavior. Additionally, due to the intrinsic self-healing characteristics of the magnetic layer the system relies solely on physical attraction where permanently magnetic particles attract each other, making the healing process automatic, reliable and rapid, without operator involvement and uninhibited by different ambient conditions. Furthermore, self-healing of the swimmer can occur at the same place numerous times as no other healing component is used but the magnetic interaction. Rational design of the geometry, alignment of the magnetic particles, and orientation of the magnetic strips enables efficient 'on-the-fly' healing behavior and reduces unsymmetrical healing. Additionally, the movement of the swimmer can bring separated pieces together despite large distances more than 10 times the size of the swimmer. This is superior to previous strategies for static magnetic healing systems which provide healing only for separations smaller than 3 mm.[23] With these attractive advantages in mind, the developed SHSs may open the door to future autonomous self-healing robotic platforms for a variety of applications.

**Results and Discussion**

**Fabrication and operation of SHSs**

Self-healing swimmers were fabricated using a versatile screen-printing technique where multiple functional layers are printed on top of each other (see Materials and Methods). The multi-step fabrication process is presented in **Figure 1A**. The main structure of the SHS features 3 layers: the conductive bottom layer, the rigid and hydrophobic middle layer and the top magnetic layer. After fabrication and release from the substrate, the SHSs are endowed with a catalytic capability by electrodepositing Pt on their tail section for which the conductive bottom layer is necessary. The rigid hydrophobic middle layer is utilized to prevent the swimmer from sinking or becoming floppy and hence ensuring strong swimming. The top magnetic layer enables the self-recombination after the damage Following this step, the SHSs can propel autonomously in the peroxide fuel solution and can recover their initial structure and original motion after suffering catastrophic damage.

A schematic of the self-healing process of a typical swimmer is shown in Figure 1B. Initially, a pristine (non-damaged) SHS is swimming autonomously until it is damaged (cut into multiple pieces). As the self-propelling tail portion of the SHS travels around and is attracted to the static body piece (upon approaching it) due to the strong magnetic interactions of the magnetic strips of the separated pieces until recombination occurs. The self-healed swimmer then restores the swimming in a manner similar to the pristine SHS. The proposed mechanism is first shown outside of solution (no motion) with a model 1 SHS (1 magnetic stripe through the middle). Figure 1C shows the non-damaged swimmer. Next, the tail is detached from the body (Figure 1D). Finally, the tail reattaches to its body confirming the strong magnetic behavior. The healed SHS is shown in Figure 1E. The time-lapse images of Figure 1F (Video S1) show a damaged SHS illustrate the reorientation and reattachment of the moving tail to its complementary static piece. At a large enough separation, the swimmer pieces do not feel each other as the magnetic attractive force is distance dependent ($F_m \propto 1/r^4$). However, as the tail gets closer, the attractive magnetic force realigns the pieces so that the built-in movement brings them together and repair can occur autonomously. Thus, healing is partially reliant on the size of the swimming vessel. Since the magnetic force falls off rapidly with distance it can only bring and align damaged portions of the SHS at distances around 50 mm. Accordingly, the SHS must come into each other's vicinity by random propulsion before the magnetic force can align and join them. Hence, increasing vessel size decreases the probability of damaged SHS pieces coming closer within 50 mm and increases the time for healing to occur. The exact distance dependence of the magnetic force and the distribution of the magnetic field around a swimmer have a profound effect upon the overall healing behavior and are discussed in detail in the following sections. As will be illustrated throughout this paper the magnetic healing strategy is very versatile and robust. For example, Figure 1G demonstrates a SHS cut into 3 pieces, named T, B and H for tail, body and head, respectively (Video S1). Additionally, the cut between the B and H sections was done at an oblique angle to test the robustness of the healing process. Following swimming of the T for few seconds it was attracted and reattached to the B, whereby the combined T+B structure restored its motion (Figure 1H). As this T+B portion continued its movement, it was attracted to the remaining H, resulting in complete recovery of the original SHS structure which continued to swim like the original pristine SHS (Figure 1I). Finally, to showcase the strength of the healing process the recovered swimmer is hung vertically in Figure 1J, demonstrating that the 3 pieces are securely attached together.

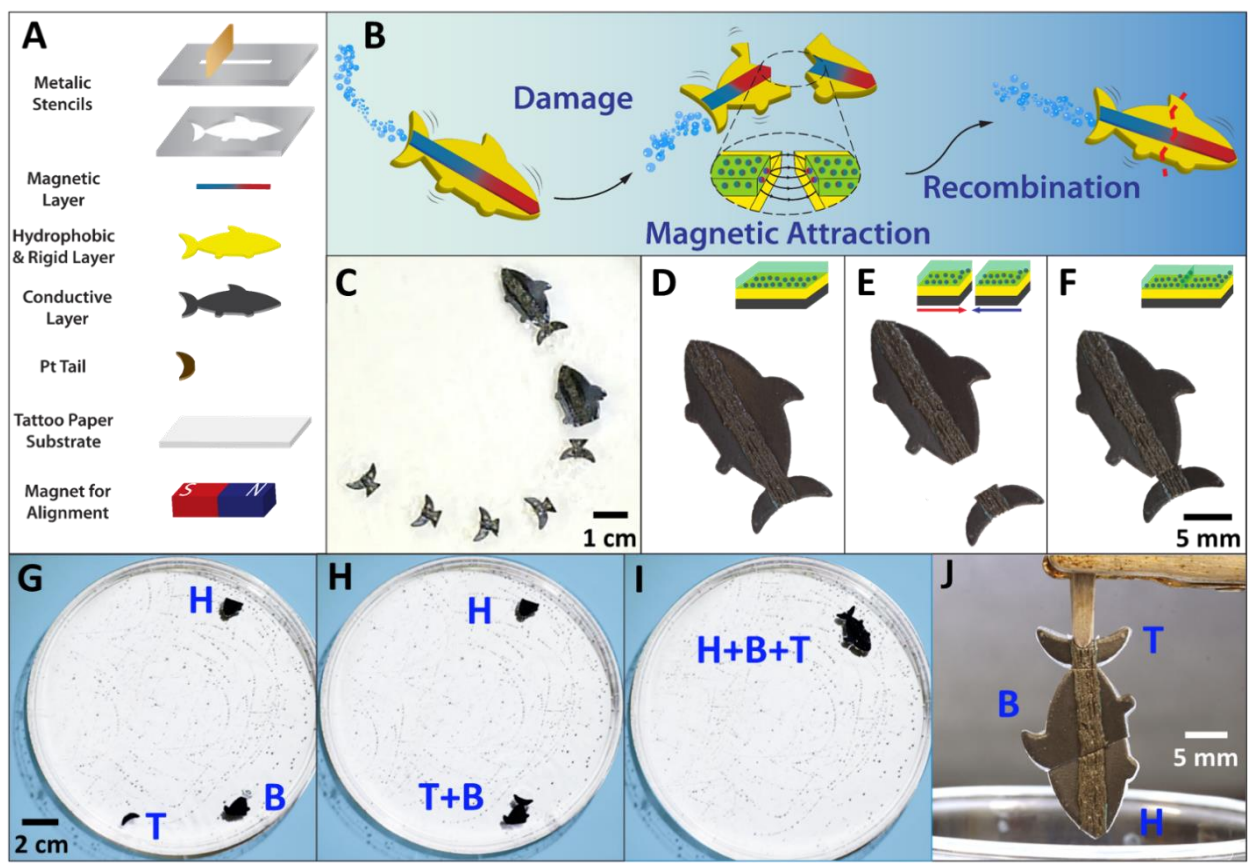

*Figure 1. Autonomous 'on-the-fly' self-healing of magnetic-based SHS. (A) Fabrication process of a SHS. (B) Diagram of a swimmer propelling in solution, experiencing extensive damage, and self-healing based on magnetic attraction. (C) Image of a pristine SHS on a dry surface. (D) Image of a swimmer damaged at the tail. (E) Image of the healed swimmer. (F) Time-lapse images of a damaged swimmer healing over the course of 1.6 s. taken from Video S1. (G) Images of a SHS damaged into three distinct pieces: head, (H), body (B) and tail (T). (H) Image of the healed tail and body. (I) Image of the completely healed swimmer. Taken from Videos S1. (J) Image of a healed swimmer suspended vertically while maintaining its integrity, showcasing the strength of the magnetic attraction of individual pieces.*

**Characterization of the magnetic behavior of SHSs**

Full realization of such remarkable 'on-the-fly' healing behavior requires a deeper understanding of the magnetic fields and forces around each piece of the damaged swimmer. The nature of the magnetic interactions within the self-healing strips is crucial for the successful reattachment of damaged swimmers. A typical model 1 swimmer with a focus on the magnetic strip is shown in **Figure 2A**. The topography of the strip was assessed *via* 3D light microscopy. As mentioned in the Materials and Methods section, the magnetic strip is printed in the presence of a strong external magnetic field in order to align the magnetic particles. The parallel grooves seen in the topographical image reconstruction attest to the alignment of the constituent magnetic particles.

Next, the magnetic strip was subjected to a magnetic hysteresis test to quantitatively assess the alignment of the magnetic particles. The normalized ($M/M_s$) hysteresis loops (Figure 2B) show typical hard magnetic behavior, with high magnetic remanence ($M_r$) and magnetic coercivity, $H_c$. The magnetic saturation ($M_s$) of the strip was ~ 87 emu/g, along with a magnetic coercivity, $H_c$, of ~2.5 kOe and a highest $M_r$ of 61.48 emu/g. Importantly, the curves reveal clear alignment of the magnetic materials in the SHSs. The red trace (D1, alignment perpendicular to the magnetic field) has a $M_r/M_s$ ratio of 0.53 while the blue trace (D2, alignment parallel to the magnetic field) exhibits a $M_r/M_s$ of 0.74. Typically, in well-aligned samples, there is a clear directional dependence of the magnetic properties, reminiscent of highly anisotropic magnetic materials.[24,25] A $M_r/M_s$ ~ 0.5 is typical of randomly oriented material while a high ($M_r/M_s > 0.5$) is indicative of well aligned magnetic material. The high $M_r/M_s$ ratio (0.74) in the D2 direction clearly shows that the magnetic particles are well aligned along the long axis of the strip.

Analysis of Video S2 provides a direct representation of the magnetic field around a swimmer or parts of a swimmer (Figure 2C), with further details found in the Supplementary Materials. Based on the orientation of the magnetic field lines it is clear that model 1 SHS magnetic strips acts as dipole permanent magnets even when separated. As the mobile tail approaches the static swimmer piece, the magnetic fields overlap until the self-healing process restores the initial structure. Finally, the overlap of the magnetic fields produces only one magnetic dipole, confirming the structural integrity of the swimmer.

Next, the magnetic field strength was verified with a handheld Gauss meter. The heat map grid of Figure 2D illustrates the concentration of magnetic flux along the body and surroundings of a model 1 SHS which has been fabricated under a strong magnetic field. Due to the magnetic alignment the SHS exhibits stronger field values compared to an unaligned SHS (model 1U) (Figure 2E). Note that the highest field values are recorded at the head and tail of the swimmer, affirming that the dispersion of magnetic particles and the alignment make the strip act as a bar magnet. Without being exposed to a magnetic field alignment, the magnetic particles within the self-healing layer of model 1U are oriented randomly and exhibit diminished magnetic behavior. Since there are moving parts involved it is very important to characterize the maximum range of the magnetic flux in the system. Figure 2F illustrates how the magnetic flux diminishes with distance for model 1 and 1U swimmers. The former displays strong magnetic flux at larger distances, realized by aligning the particles during fabrication. This was also verified by model 1U swimmer in solution (Figure S1, Video S3). Despite the approach of the active tail to the passive piece and the physical contact of the 2 pieces, there is no lasting attraction and hence no structural recovery, confirming the crucial importance of magnetic alignment during fabrication.

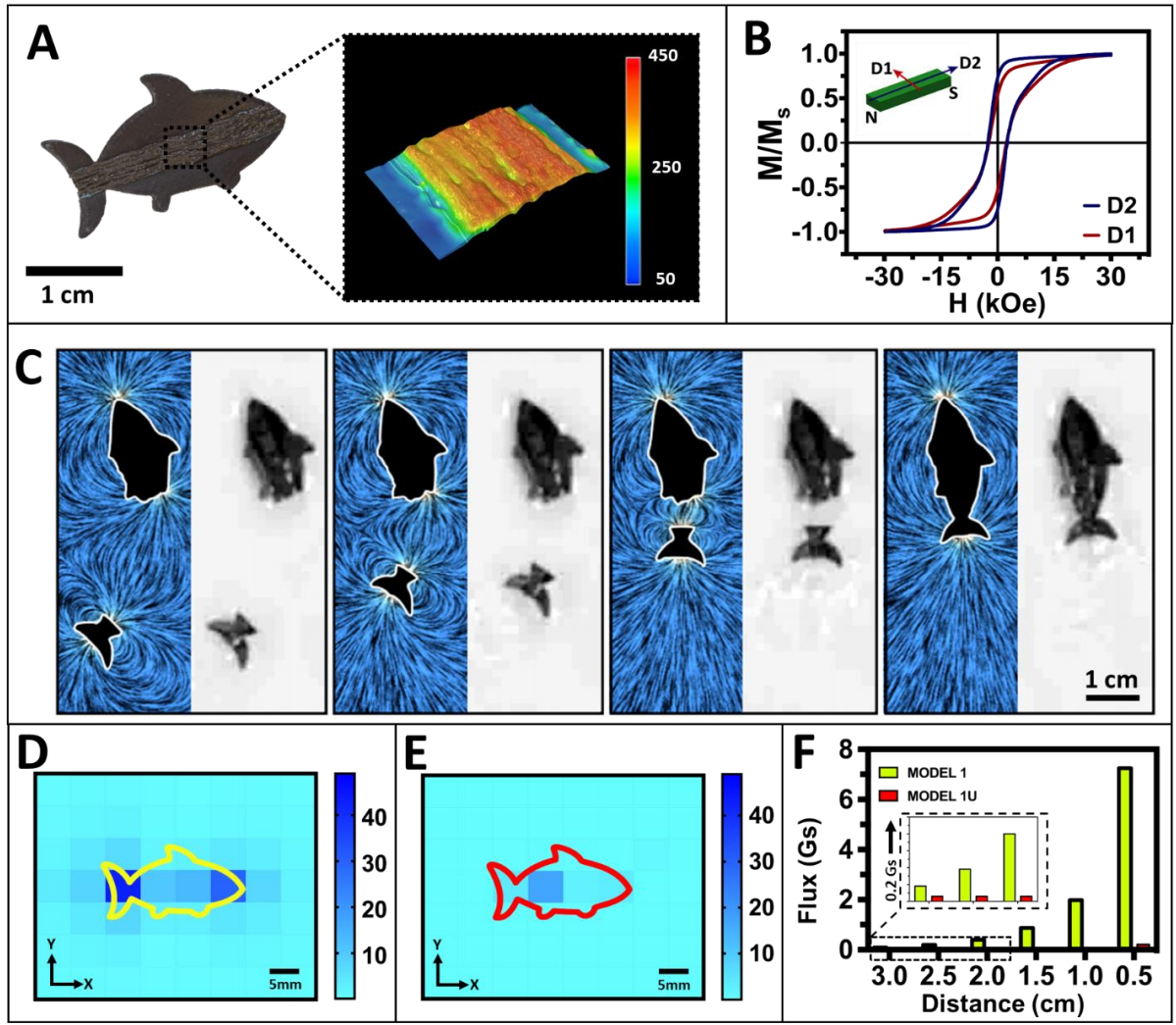

*Figure 2. Assessment of the magnetic behavior of SHSs. (A) Image of a model 1 SHS. Inset represents topographical image of the magnetic strip of the SHS. Heat map is in μm. (B) Magnetic hysteresis loop of a magnetic strip perpendicular (D1, red) and parallel to the alignment axis (D2, blue). (C) Representation of the magnetic field around a SHS during a healing event. Taken from Video S2. (D) Grid of magnetic field measurement around a model 1 SHS. Heat map is in Gauss. (E) Grid of magnetic field measurement around a model 1U SHS. Heat map is in Gauss. (F) Magnetic flux as a function of separation distance for model 1 and model 1U SHSs. Inset shows a magnification of the region from 3 to 2 cm.*

**Effect of the damage position on healing behavior and efficiency**

To consider the role of fluid dynamics upon the efficiency of such 'on-the-fly' healing an experiment where a swimmer was immobilized (fixed) to the surface but allowed to react with the fuel and produce bubble thrust. Tracer particles were introduced to capture the flow of the liquid around the confined swimmer. For a typical SHS of length scale ~ 1 cm moving with speed of ~ 10 cm/s in a fluid of density ~1000 kg/m$^3$ and viscosity ~ 0.001 Pa·s the Reynolds number is about 1000. Far from the low Reynolds number regime inertia can have a significant influence on swimmer motion. Figure S2 shows the flow field around a fixed swimmer. The flow pattern near the tail differs significantly from the flow pattern in a steady flow (around the rest of the swimmer). The bubbles leave the tail and apply momentum to the fluid. Upon bursting on the surface, the bubbles push the fluid away from the tail and provide the propulsion force. Considering this fluid dynamic behavior, we expect self-healing to be dictated not only by the magnetic attraction but also by the dynamics of the system propulsion.

Next, we investigated the self-healing capabilities as a function of the damage location using model 1 SHSs. We studied three different cases (**Figure 3A**). Case i features a cut 7 mm away from the tail end of the swimmer. Case ii features a cut exactly in the middle of the swimmer, while case iii corresponds to damage 6 mm away from the front of the SHS. Additionally, to evaluate the healing effectiveness of the model 1 SHS based on the damage position we define a healing efficiency (HE) as the number of times the swimmer has successfully and autonomously healed, retaining both the pristine structure and initial propulsion, out of the all healing events.

$$(HE = \frac{Healing\ events\ with\ recovered\ propulsion\ and\ correct\ structure}{All\ recombination\ events} * 100\%).$$

We can go further and calculate the overall healing efficiency (OHE) which also takes into account the cases when the healing is imperfect (mismatching), while the swimmer retains the propulsion capability.

$$(OHE = \frac{Healing\ events\ with\ recovered\ propulsion\ and\ mismatched\ structure}{All\ recombination\ events} * 100\%)$$

Finally, the remaining healing events feature a reattachment with large structural mismatch and loss of the autonomous motion (SHS is spinning or stuck on the wall). A real example of case i SHS is demonstrated in Figure 3B (Video S4). In this case, the passive body is large relative to the mobile tail and does not rotate to match with the self-propelling tail because of its size and weight. On the other hand, the tail moves rapidly, fixing itself to the body and reattaching easily. Case i exhibits a HE of 88% with an OHE of 94%. For case ii, both pieces exhibit some rotation and alignment as their respective sizes are very similar (Figure 3C, Video S4). Case ii demonstrates a HE and OHE of 86%, demonstrating that the size of the passive and active portions of the swimmer affect the healing process. In case iii, the smaller head has a higher propensity to change its direction and align to the magnetic field of the larger propelling tail-containing body (Fig 3D, Video S4). Yet, at the same time the force of the catalytic propulsion overcomes the magnetic force, affecting the healing capability of the body, leading to diminished healing capability, indicated by a HE of 58% and OHE of 71%. Furthermore, we compared these results with predictions of the magnetic force between the magnetic strips of a damaged swimmer from an analytical model (additional details in the Supplementary Materials). As expected, the force falls off rapidly with separation distance according to $\tilde{F} \sim d^{-4}$. More interestingly, the magnetic force depends on the position of the cut (x). With smaller x (larger size difference between the two damaged pieces) the force is smaller, suggesting lower healing ability. However, as noted above, the healing behavior depends not only on the magnetic force but also upon the propulsion behavior and relative size of the swimmers.

To investigate this further, we performed simulations of the magnetic force and torque of the swimmers in many orientations and distances away from each other (Figure S3). Here, we refrained from adding the dynamic motion of the tail sections to remove temporal changes and elucidate the aligning nature of the magnetic interaction. We chose to look at torque because this quantity is directly responsible for the rotation and alignment of magnetic strips at various orientations. At small distances (5 mm) the torque increases as the rotation of the mobile strip grows more perpendicular to the stationary one. This is expected as the more misaligned (perpendicular) the strips are the larger the torque will be to bring them back to alignment. However, upon increasing the distance between the strips (20 and 50 mm) the torque assumes a maximum before the perpendicular case, suggesting that the alignment is preferred at smaller rotation angles (θ). Additionally, when the approach angle is large (φ) the torque is higher for virtually all cases, suggesting that the susceptibility for alignment is stronger on the periphery of the half-disk described by φ. Conceptually, it makes sense that the torque is low at low angles of φ and θ as the strips are already nearly aligned, suggesting high probability of healing. With more misalignment the torque is higher, meaning an induced rotation of the swimmer will occur to align them and produce self-healing. Finally, at large values of θ, φ and D the torque will surpass a maximum and decrease, suggesting that beyond these points self-healing is more unlikely. Overall, the data of Figure S3 suggest that the magnetic torque will most likely align damaged pieces at all orientations at short distances while at large distances the orientations which will lead to healing become restricted.

Additionally, we investigated the effect of the SHS speed on the self-healing behavior. For this purpose, SHSs with different speeds prepared by electroplating different amounts of Pt (using 20 or 120 s deposition; 60 s used in all other experiments). Figure 3F presents an SHS with 20 s Pt deposited tail. The trajectory exhibits the motion of the pristine (a) and healed (b) cases over a 4s period, illustrating a very similar propulsion behavior after healing. For the much faster swimmer (based on the 120 s Pt plating), the pristine (a) and healed (b) trajectories show little variation (Figure 3G). Finally, we compared the speed of all three Pt deposition examples for the pristine and healed cases. A very small difference between the speed of the pristine and healed swimmers is observed for these different catalytic swimmers, with the speed of healed swimmers being slightly lower compared to the pristine ones (Figure 3H; a vs b).

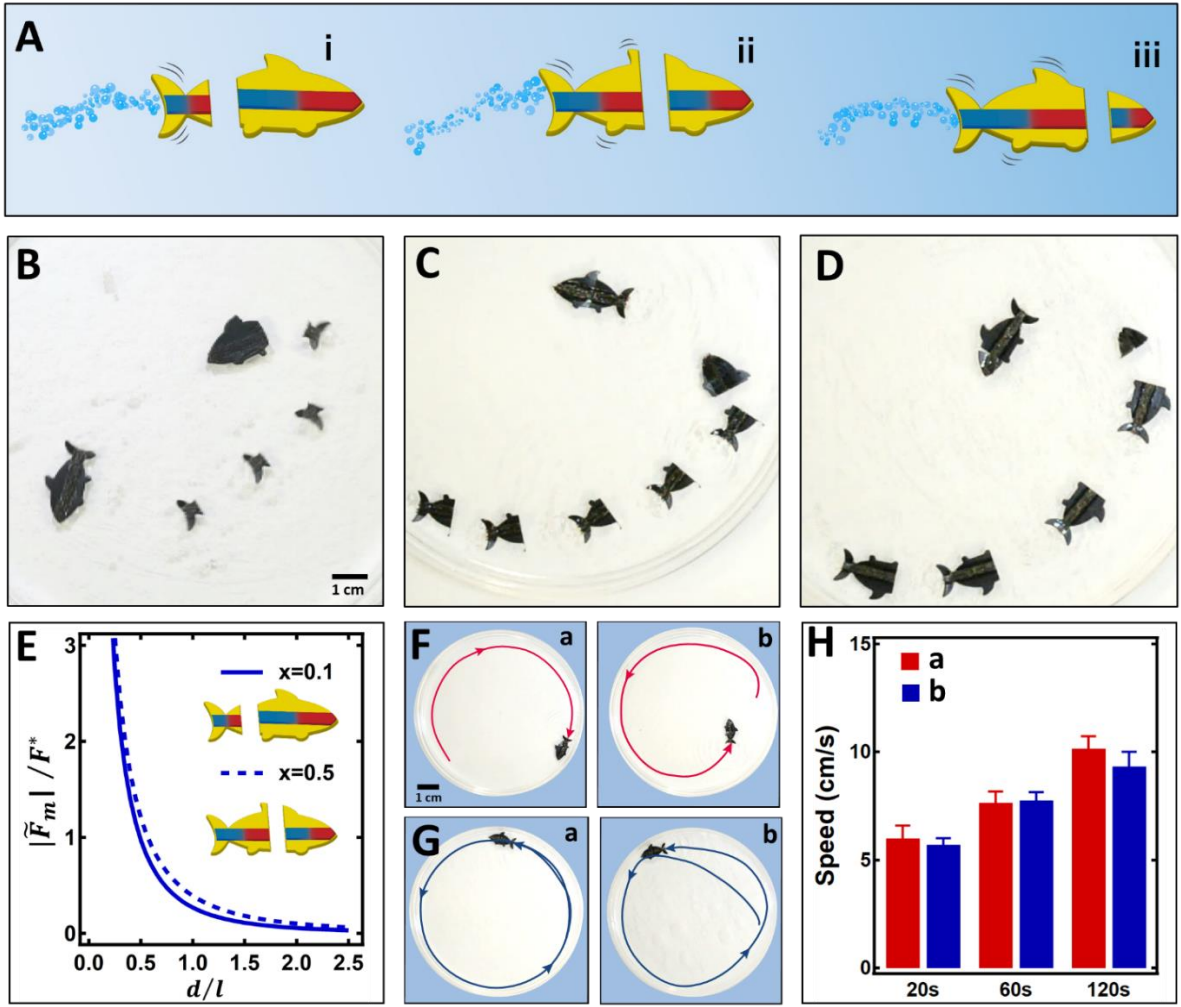

*Figure 3. Influence of the damage position upon the propulsion behavior. A) Diagram of some possible damage positions along the SHS: damage close to the tail (case i), halfway through the structure (case ii), and close to the head (case iii). B, C, D) Time-lapse image of a model 1 SHS pertaining to cases i, ii, and iii, respectively. Taken from Video S4. E) Plot of the normalized magnetic force as function of the dimensionless separation distance for two different cut positions (x=0.1 and 0.5) from an analytical model. (F) Propulsion of a pristine (a) and healed (b) SHS with 20 s deposited Pt on the tail. Trajectory shows motion over a 4 period. (G) Propulsion of a pristine (a) and healed (b) SHS with a 120 s Pt deposition on the tail. Trajectory shows the motion over a 4 s duration. (H) Effect of the Pt deposition time upon the speed of the pristine (red) and healed (blue) SHSs.*

**Effect of magnetic strip configuration on healing behavior and efficiency**

Assessing the healing behavior of different magnetic configurations is an important step towards optimizing the healing process and configuring this strategy for various applications. **Figure 4A** showcases the schematic (a) of a model 1 SHS. Additionally, we provide a visual representation of the magnetic field lines around each swimmer to aid in understanding the healing behavior (b) and an image of the swimmer (c). Zoom-in images of the recombined swimmers are displayed in Figure S4. Furthermore, we show images of the self-healing process. For a model 1 (damaged in the middle) the propelling tail easily finds its complementary passive section, reattaches and with the healed swimmer continues to move similar to the pristine version (c) (Video S5). A similar healing behavior is observed for the two-strip (model 2) swimmer (Figure 4B, Video S5). Compared to model 1, model 2 SHSs demonstrate a HE of 49% and OHE of 69%. Thus, the two-strip configuration is not as effective as model 1. The larger difference between the HE and OHE indicates that there is a large number of imperfect healing events. This can be attributed to the larger surface that multiple strips cover. Using the same magnetic alignment, model 3 also presented effective healing after being damaged in the middle (Figure 4C, Video S5), exhibiting a HE of 41% and OHE of 63%. For model 3, the magnetic healing surface is spread out over a larger area preventing accurate healing, but still retaining a high percentage of imperfect healing events. At the same time, the large area of magnetically-active material may cause the attachment of passive pieces to multiple areas of the swimmer besides the specific damaged area. This is supported by the representation of the magnetic field around the swimmer models 2 and 3. (b). Figure S5 presents several examples of imperfect healing (top row: imperfect healing considered for OHE; bottom row: unsuccessful healing). For example, SHSs can exhibits reattachment that is offset or imperfect, noticed particularly for models 2 and 3, owing to the large area of active magnetic material along the cut. Another improper healing even occurs when the damaged pieces heal in an incorrect order, such as the cases for models 1 and 2. In these examples, the propelling tail may randomly reattach first to the head (instead of the body) resulting in undesirable configurations.

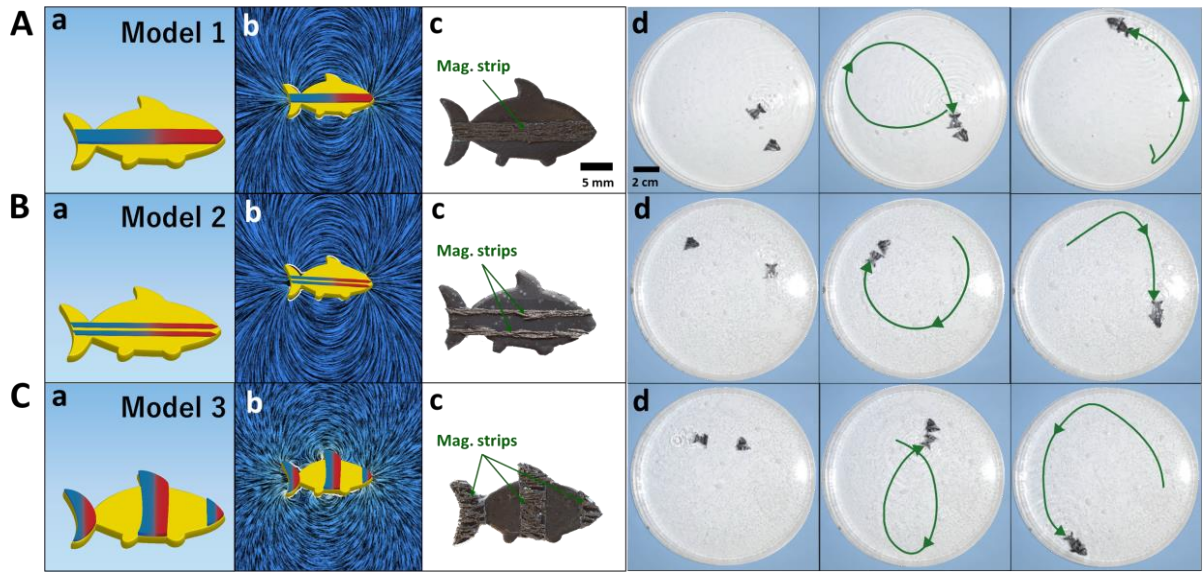

*Figure 4. Healing behavior of different magnetic strip configurations. (A) Model 1 SHS: schematic (a), visual representation of the magnetic field lines (b), zoom-in image (c), and images of damaged, approaching and healed states (d). (B) Model 2 SHS: schematic (a), visual representation of the magnetic field lines (b), zoom-in image (c), and images of damaged, approaching and healed states (d). (C) Model 3 SHS: schematic (a), visual representation of the magnetic field lines (b), zoom-in image (c), and images of damaged, approaching and healed states (d). Taken from Video S5.*

**Conclusions**

The present work demonstrates the first example of a functional small swimmer, capable of responding to damage, and restoring its structure autonomously and 'on-the-fly'. The SHSs are easy to fabricate without resorting to complex synthetic efforts and offer many benefits for future robotic endeavors. These include robust and facile healing, ability to self-repair repeatedly in the same location in the absence of external stimuli or user input, independence of environmental conditions, and versatility in design. Furthermore, we evaluate the healing performance employing different geometries of the magnetic layer showcasing that the magnetic strip concept is robust and adaptable to many configurations or material systems, depending on the specific application. While the presented swimmer moves on the surface of water, this concept can be readily applied to 3D swimmers moving in bulk fluids. For example, the incorporation of a magnetic spine (analogous to our magnetic strip design) could provide self-repair capabilities of such innovative platforms. Such attractive and versatile healing behavior paves the way for a new class of robots that can regain their functionality after suffering extreme mechanical damage.

Despite these distinct advantages and potential benefits, several challenges must be addressed before further practical use. For example, the healing process is still prone to imperfect healing events. For the next generation smart robots, programmable and intelligent self-healing strategies which bypass misaligned or out-of-order recombination must be developed. We envision that adding more complexity to these swimming platforms, e.g., by incorporating stimuli-responsive materials, will add a basic feedback capability, and impart adaptability to their surrounding environment. As seen above, the magnetic interaction can be weak over large distances, hindering the healing in large reservoirs and requiring strategies for actively directing swimmers towards a target location or preconcentrating them. Additionally, non-toxic fuels must be developed to replace the peroxide fuel used in this proof-of-concept study.

**Materials and Methods**

*Fabrication of SHSs:* The fabrication process was comprised of printing various functional inks. The components of the swimmer body and tail were designed in AutoCAD (Autodesk, San Rafael, CA) and used as patterns in the 12″ × 12″ stainless steel stencils (Metal Etch Services, San Marcos, CA). A temporary tattoo paper sheet (Papilio, HPS LLC, Rhome, TX), precoated with a water-soluble adhesive, served as the substrate for the printing process. A typical SHS included the printing of three layers using an MPM SPM semi-automatic screen printer (Speed-line Technologies). First, the conductive layer from graphite ink (E3449, Ercon, Inc., Wareham, MA) was printed in the shape of the swimmer and cured at 60 °C in an oven. This conductive layer enabled subsequent electrodeposition of Pt on the tail to enable propulsion. Next, three consecutive layers of a hydrophobic, rigid layer was printed. The hydrophobic layer consists of 30 wt% solution of polystyrene-polymethylmethacrylate copolymer (Aldrich, St. Louis, MO) in toluene with 5 wt% aerogel particles (<20 μm, Jios Chemicals, South Korea). Lastly, magnetic $Nd_2Fe_{14}B$ powder was obtained following previously reported protocols.[23] Fabrication of the magnetic ink consisted of preparing a 30 wt% dispersion of magnetic powder in polymeric insulator ink (Dupont 5036, Dupont, Wilmington, DE). In order to align the magnetic particles, the SHS was exposed to a strong magnetic field using a commercial magnet (CMS Magnetics).

*Deposition of Catalytic Pt:* Swimmers were removed from the tattoo paper substrates by soaking them in water for 30 s and then sliding the SHS off the edge of the substrate. Next, SHSs were subjected to an electrochemical deposition of a Pt layer on their tail using a commercial Pt plating solution (Technic Inc., Anaheim, CA) in a standard three-electrode setup. The carbon tail of the swimmer served as the working electrode, while a Pt wire and Ag/AgCl (1 M KCl) served as the counter and reference electrodes, respectively. The galvanostatic deposition of Pt was performed on a μ-Autolab potentiostat (Metrohm Autolab B. V., Netherlands) at −2 mA for 20, 60 or 120 s.

*Swimming and Self-Healing:* Self-healing swimmers were tested inside a 150 mm diameter Petri dish filled with 70 mL of 15% $H_2O_2$ solution. A camera mounted on a tripod was placed above the Petri dish to record videos of the swimming and healing processes. Damage to the SHSs was introduced with a blade and the SHS pieces were placed back into the solution in opposite sides of the Petri dish and left to autonomously heal.

*Magnetic Hysteresis and Field Distribution Measurement:* A Quantum Design Versalab with a VSM attachment was used to measure the magnetic hysteresis of the SHSs. Printed magnetic strips were measured at room temperature (300K) with an applied field sweeping from -30 to 30 kOe. Each strip was measured in two orientations which were perpendicular to each other (hereafter designated as D1 and D2). The direction D1 denotes the orientation perpendicular to the magnetic alignment inside the strip while D2 denotes the orientation parallel to the magnetic alignment in the strip.

Measurements of the actual field distribution around an SHS were performed with a Lakeshore Model 425 Gauss meter and probe. An area of 35 × 40 mm was separated into a 7 × 8 grid with an individual box size of 5 × 5 mm. The SHS was placed in the middle of the grid and at each grid box the value of the magnetic field was measured to produce a field distribution map.

*Equipment:* Videos were taken on a Nikon D7000 camera with a Micro Nikon 40 mm lens mounted on a tripod. The resulting videos were analyzed with Nikon Elements AR 3.2 tracking software.

*Statistical Analysis:* All quantitative values were presented as means ± SD. All experiments were performed for at least three independent repeats.

**Supporting Information**

Additional material includes examples of unsuccessful healing in unaligned strips, Figure 3 model explanation, simulation of magnetic torque with varying distance and angle of approach, zoom-ins of healed swimmers from Figure 4, examples of misaligned healing of model 1,2 and 3 swimmers, and explanation of supporting videos.

**Acknowledgements**

modeling, respectively. All authors assisted in data analysis and the manuscript preparation. Additionally, the authors thank Xiaolong Lu for fruitful discussions. The authors acknowledge Fernando Soto for assisting with the 3D light microscopy. This work was supported by the Defense Threat Reduction Agency Joint Science and Technology Office for Chemical and Biological Defense (grant no. HDTRA1-14-1-0064). EK acknowledges the Charles Lee Powell Foundation for support. C.S. acknowledges support from UC-MEXUS-CONACYT.

†These authors contributed equally to this work.

**Conflict of Interst**

The authors declare that no competing interests.

# Supporting Information

**Self-Healing Small-Scale Swimmers**

*Emil Karshalev[1] †, Cristian Silva-Lopez[1] †, Kyle Chan[2], Jieming Yan[1], Elodie Sandraz[1], Mathieu Gallot[1], Amir Nourhani[1], Javier Garay[2] and Joseph Wang[1] ***

[1] Department of NanoEngineering, University of California San Diego, La Jolla, CA 92093, USA.
[2] Department of Mechanical and Aerospace Engineering, University of California San Diego, La Jolla, CA 92093, USA.

*Corresponding author. Email: josephwang@eng.uscd.edu.

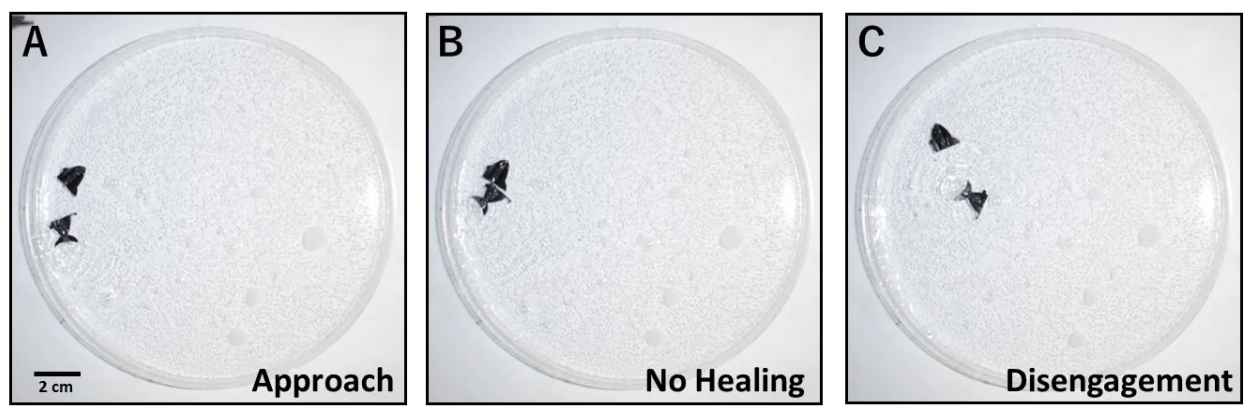

*Figure S1. Unsuccessful healing of a model 1U SHS. (A) Two components of the swimmer approach each other. (B) The pieces touch but no reattachment occurs. (C) The self-propelling piece moves away from the static piece.*

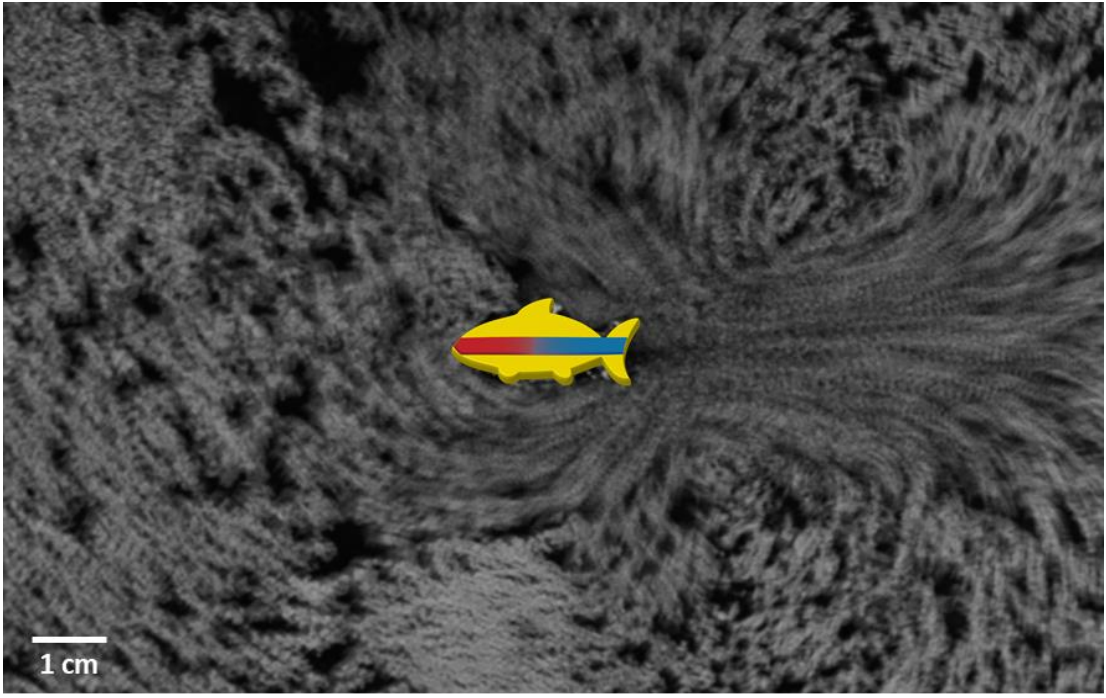

*Figure S2. Flow pattern around a SHS.*

To gain a better understanding of the magnetic forces and how they vary with distance and magnetic strip separation we developed an analytical model (corresponds to Fig 3E). The separated strips behave as two permanent magnets with lengths $L_1$ and $L_2$ and $L=L_1 + L_2$. We define the fraction $x=L_1/L$ as the fraction of the strip that contains all or part of the tail. A magnetic strip on a self-healing fish has a typical length L=20 mm, a width of l=2.5 mm and thickness of t=0.270 mm. Using the width as a length scale, $t/l \sim \Theta\,(10^{-1})$ we can approximate the magnetic bars as two-dimensional objects. Using the method of magnetic pole strength and approximating the bars as 2D rectangles we obtain the magnetic force between the strips (Equation 1)

$$F_m(L_1, L_2) = f(L_1 + d) + f(L_2 + d) - f(L + d) - f(d) \qquad (1)$$

with $f(x) = \frac{\mu_0}{2\pi}(tM)^2\left[\sqrt{1 + (l/x)^2} - 1\right]$ where $\mu_0 = 4\pi \times 10^{-7}$ H/m is the vacuum permeability, t is the thickness of the magnetic strip, l is the width of the magnetic strip, and M is the magnetization density. Defining the force scale as $F^* = \left[\frac{\mu_0}{2\pi}(tM)^2\right]$, the dimensionless force between the bars $\tilde{F}_m = F_m/F^*$ is plotted vs the dimensionless distance $(d/l)$ for different cut positions (x) and an aspect ratios of the magnetic strips $L/l = 8$ as it corresponds to a Model 1 swimmer (where d is the separation distance between the magnetic strips). As the plots shows, the maximum attraction force happens when the magnetic strip parts into two equal segments (x=0.5). Also, the force increases upon increasing the swimmer size, and for a very long swimmer (L≫l) we reach the asymptotic value of $\tilde{F} \sim 1 - \sqrt{1 + (l/d)^2} \sim d^{-4}$ which is the magnetic force between two point dipoles. The force between the magnetic strips drop rapidly as a function of distance between the nearest poles. Thus, one of the main roles of magnetic attraction in the self-healing process is the alignment of the orientation of the magnetic strips to facilitate 'on-the-fly' healing.

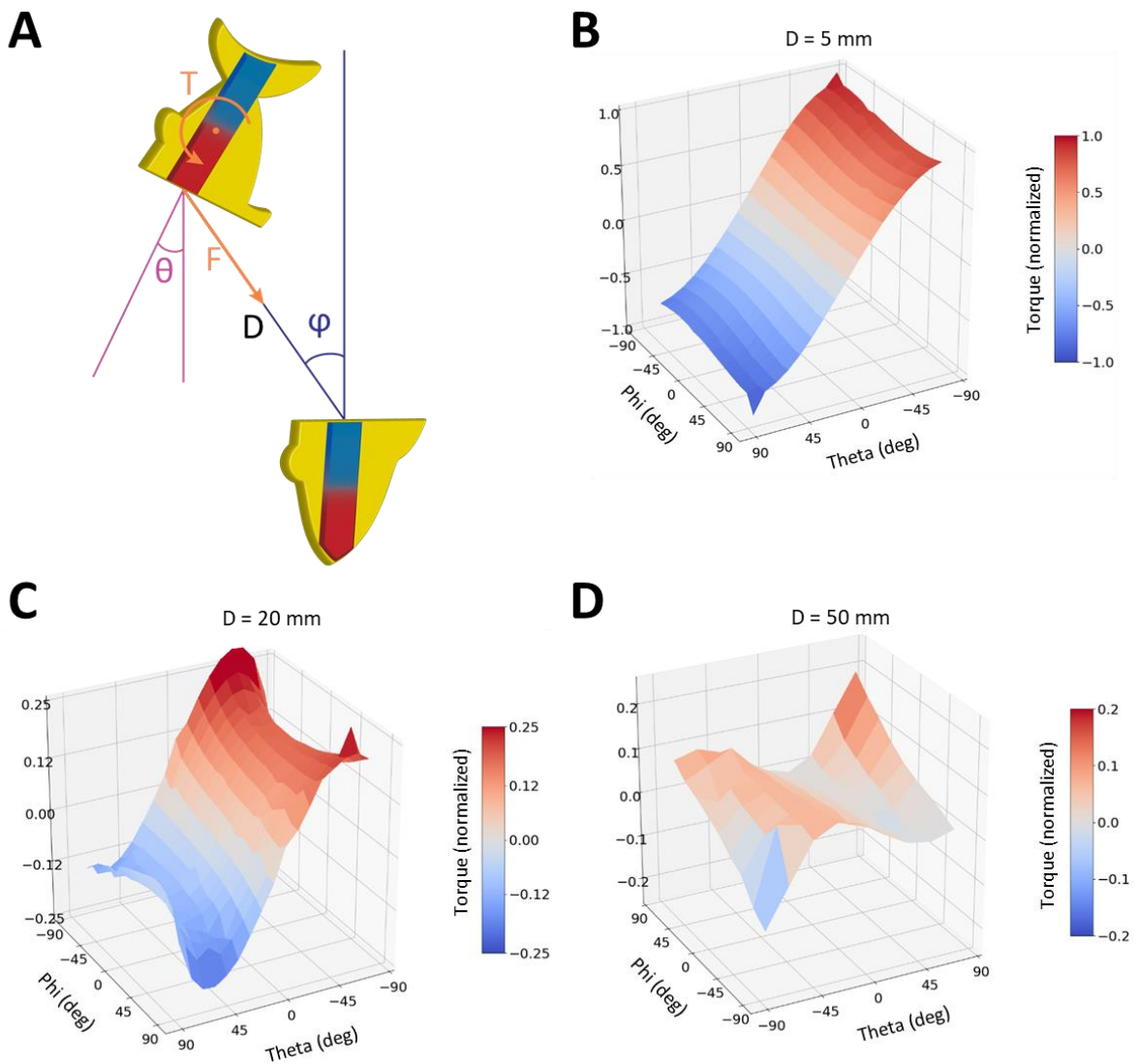

*Figure S3. Simulation of the magnetic torque towards alignment in different scenarios. (A) Schematic representation of the orientations a damaged swimmer can assume depending on*

*the rotation angle (θ), approach angle (φ), and distance (D). Normalized torque as a function of the rotation and approach angles for small (B), medium (C), and large (D) distances.*

COMSOL Multiphysics 5.2 software with the AC/DC>Magnetic Fields, no Currents (mfnc) module was chosen along with a two-dimensional Stationary Study. The study was focusing on the magnetic interactions between the two magnetic strips comprising the swimmers in order to deduce trends of the magnetic forces and of magnetic torques for various positions of the stripes (varying distance and orientation between them). The two magnetic strips were modeled as 2 identical rectangles made out of $Nd_2Fe_{14}B$, in a surrounding disk of air. The first strip had a fixed position in the center of the geometry while the latter one was positioned in a half-disk around the first one where their poles could face each other Figure S3A. Precaution should be taken using COMSOL for such model and one should consider the effects of boundary conditions, the appearance of numerical noise and the impact of the meshing. The dimensions of the circle of air were set to 100 times larger than the length of the stipes to prevent boundary conditions from affecting the magnetic field. The meshing has been tuned to be extremely fine on the stripes, fine in a 2-meters circle surrounding the swimmers and coarse over 2 meters. Finally, as the theory assures that the magnetic forces applied from one magnet to another are equal, we computed the numerical noise as the relative difference between the two forces and tuned the meshing to make sure that this noise would not exceed 5% in the range of the positions we used. The relative magnetic permeability of the stripes has been set to 1.05 and the magnetization to 468 kA/m (Figure 2B) along their length. The magnetic force has been computed from the end of the moving magnet and the magnetic torque has been computed on the center of mass of the moving strip. Positions have been chosen to describe the behavior of the mobile magnet from -80° to + 80° around the immobilized magnet (φ), from -80° to + 80° around its center of mass (θ) and the distance between them from 5 mm to 50 mm.

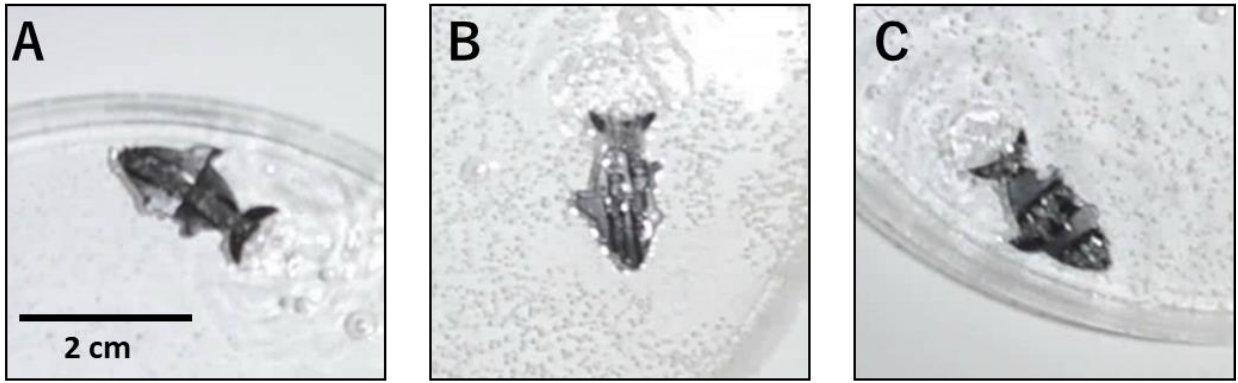

*Figure S4. Zoom-ins of self-healed model 1 (A), model 2 (B) and model 3 (C) swimmers, respectively, from Figure 4.*

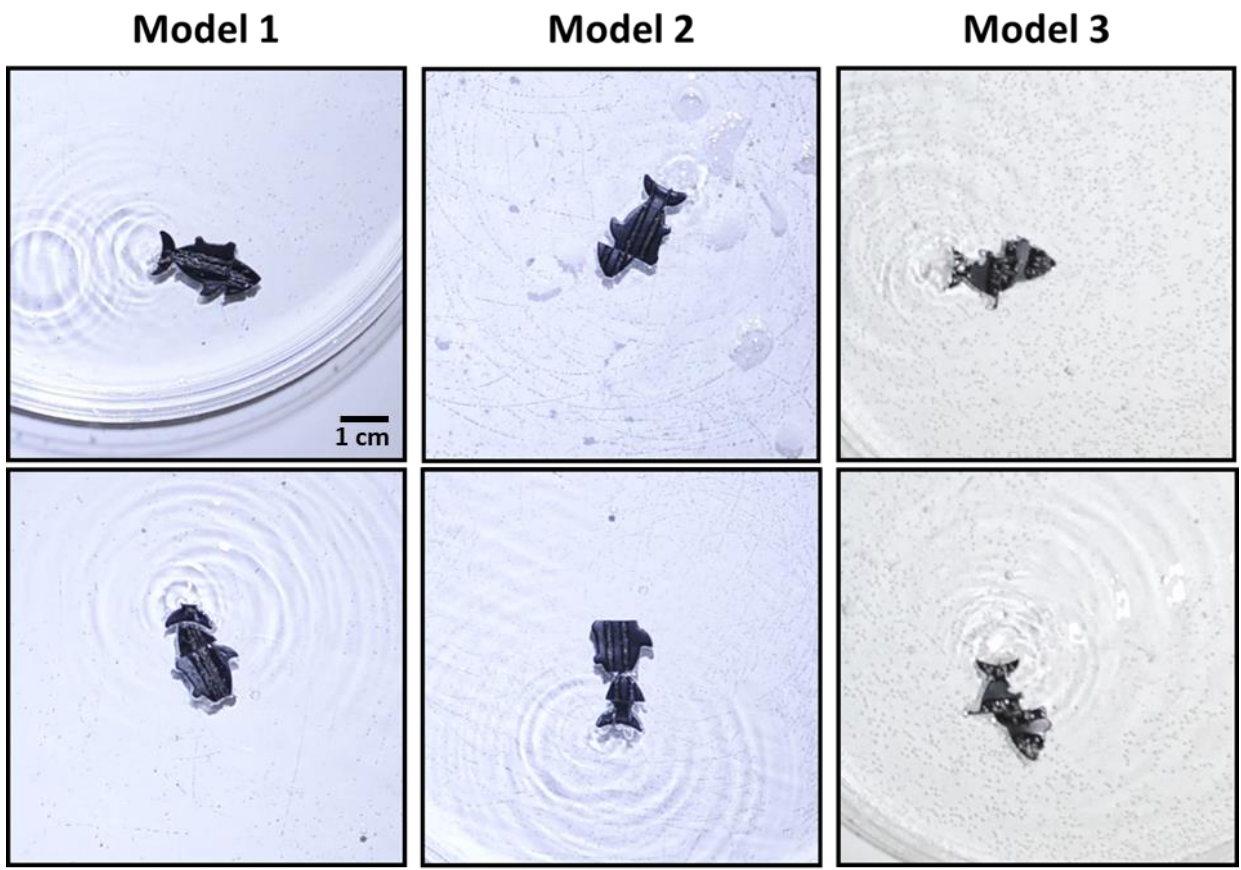

*Figure S5. **Misaligned healing.** Examples of misaligned healing of model 1, 2 and 3 SHSs featuring misalignment of the magnetic strips and incorrect order of healing for the 2 cut cases.*

Video S1. Self-Healing Swimmers

Video S2. Magnetic Field Distribution During Self-Healing

Video S3. Unsuccessful Healing of Model 1U Swimmers

Video S4. Effect of Cut Position on Self-Healing

Video S5. Effect of Strip Geometry on Healing